\documentclass[preprint,12pt]{elsarticle}

\usepackage{graphicx}
\usepackage{epsfig}
\input{epsf.sty}
\usepackage{amssymb}

\newcommand{\beq}{\begin{equation}}
\newcommand{\eeq}{\end{equation}}
\newcommand{\ba}{\begin{array}}
\newcommand{\ea}{\end{array}}
\newcommand{\beqa}{\begin{eqnarray}}
\newcommand{\eeqa}{\end{eqnarray}}
\newcommand{\bd}[1]{ \mbox{\boldmath $#1$}  }
\newcommand{\oh}{\frac{1}{2}}

\newcommand{\alp}{\alpha}
\newcommand{\lam}{\lambda}
\newcommand{\nn}{\nonumber}

\newcommand{\APNY}[1]{Ann. Phys.(N.Y.){\bf {#1}}}

\newcommand{\NPA}[1]{Nucl. Phys.{\bf A{#1}}}
\newcommand{\PLB}[1]{Phys. Lett.{\bf B{#1}}}
\newcommand{\PRev}[1]{Phys. Rev.{\bf {#1}}}
\newcommand{\PRA}[1]{Phys. Rev.{\bf A{#1}}}
\newcommand{\PRB}[1]{Phys. Rev.{\bf B{#1}}}
\newcommand{\PRC}[1]{Phys. Rev.{\bf C{#1}}}

\newcommand{\PRL}[1]{Phys. Rev. Lett.{\bf {#1}}}

\newcommand{\PPNP}[1]{Prog. Part. Nucl. Phys.{\bf {#1}}}

\journal{Physics Letters B}

\begin{document}

\begin{frontmatter}




\title{\large{\bf Saturation and Condensate Fraction Reduction of 
Cold Alpha Matter}}


\author{F. Carstoiu} 
\author{\c S. Mi\c sicu\fnref{e-mail}}
\ead{misicu@theory.nipne.ro}
\ead[url]{http://www.theory.nipne.ro/~misicu/}
\address{NINPE-HH, Division of Theoretical Physics,
Bucharest-Magurele, POB MG-6, Romania}

\begin{abstract}
The ground state energy of ideal $\alp$-matter at $T=0$  
is analyzed within the framework of variational theory of Bose quantum liquids. 
Calculations are done for three local $\alp-\alp$ potentials 
with positive volume integrals and two-body correlation 
functions obtained from the
Pandharipande-Bethe equation. The  energy per particle of $\alp$ matter 
is evaluated in the cluster expansion formalism up to 
four-body diagrams, and using the HNC/0 and HNC/4 approximation 
for a Bose liquid. At low densities the two methods predict
similar EOS whereas at higher densities they are sensitively different, 
the HNC approximation providing saturation at lower density, bellow the
saturation value of nuclear matter. Inclusion of higher order terms 
in the cluster expansion of the condensate fraction is leading to a 
stronger 
depletion of the alpha condensate with the density compared to the two-body 
approximation prediction. 
\end{abstract}


\end{frontmatter}


\section{Introduction}
\label{sect1}

Renewed interest in the properties of $\alpha$ matter is manifest in the 
literature especially in connection with $\alpha$-particle 
Bose-Einstein condensation (BEC) in $\alpha$-like nuclei (see \cite{Sch07}
and references therein). 
Calculations reported in this reference are pointing to the existence 
of a Bose-Einstein condensate of $\alpha$-particles at low densities. 
It was also noted by these authors that with increasing density the 
condensate fraction is reduced such that at 
density corresponding to the saturation of nuclear matter
($\approx$ 0.04 $\alpha$ particles per fm$^{3}$), 
the condensate fraction is reduced to roughly one half.
The estimation of the condensate fraction was done in the lowest 
approximation, i.e. the radial distribution function (RDF) is approximated by 
the square of the two-body correlation function (CFN), and therefore it is 
less justified for higher densities. There is howewer an old estimation by 
Clark and Johnson \cite{Cla80} for three values around the saturation density 
of nuclear matter using the hypernetted chain approximation in the lowest 
order (HNC/0), i.e. taking into acount only nodal diagrams in the infinite 
density expansion of the RDF. It provides a severe reduction of the 
condensate fraction ($\approx$15\%)
compared to the lowest-order cluster expansion at the same density. On the 
other hand calculations of the cold $\alpha$ matter equation of state (EOS) 
reported by the same authors within the HNC/0 approximation and using the 
soft core  $\alpha-\alpha$ potential of Ali and Bodmer \cite{AB66} are 
predicting the 
saturation point at a high density ($\rho_{\alpha}\approx$ 0.085 $\alpha$ 
particles per fm$^{3}$). 
These benchmark calculations of the $\alpha$ matter EOS were 
very recently compared to results obtained in the frame of the 
scalar $\phi^6$ effective field theory with negative quartic and 
positive sextic interactions, to simulate the attractive character 
at long distances and repulsive at short distances, and found to 
be in a very good agreement \cite{sedrak06}.   
Though no estimations of the condensate fraction are provided for this 
high density saturation point, from the estimation made at lower 
densities, as quoted above, we expect a stronger depletion of the BEC. One is then 
confronted with the problem that at low densities the $\alpha$ matter condensate is 
far from equilibrum, whereas at the saturation point the condensate fraction is small. 

It was advocated that beyond a critical density ($\rho_\alp^*\approx$0.03 
nucleons per fm$^{3}$), due to the strong overlap of the wave-functions and the 
unavoidable action of the Pauli principle, a total extinction of the 
$\alpha$ structure 
should occur \cite{roepke99}. 
The phenomenological $\alp-\alp$ potentials used in the past
 are systematically predicting saturation of $\alp$ matter at densities considerably larger than this critical densities.
It would then be important to establish if the saturation of the $\alpha$ matter takes place below this critical density if one employs other types of potentials
that incorporate more microscopic input.

The aim of this letter is to analyse the $\alpha$ matter EOS over a wide range of 
densities and try to find the optimal CFN which reflects the interplay between 
the strong short-range and the long-range correlations 
that ultimately would lead to saturation. Gaining 
insight in the saturation properties of $\alpha$ matter could also shed light 
on the condensate fraction reduction issue. 

In what follows the g.s. energy of an infinite system of 
neutral $\alpha$-particles interacting via two-body forces is calculated within the 
variational theory  of Bose liquids. As input to the energy calculation we use  a 
prescriptions for the CFN obtained by extremizing the energy functional in the 
two-body cluster approximation. The g.s. energy is 
then calculated via the cluster expansion adding the three-body and four-body
correlations  and with the HNC method that is more reliable in the 
high density sector. 

\section{Two-Body Potential}
\label{sect2}

The $\alpha-\alpha$ potential is necessarily nonlocal due to the short range
repulsion between $\alpha$ particles coming from exclusion effects. Phenomenological potentials obtained by inversion suffer of some 
difficulties which have not been yet fully resolved,
mainly due to the limited range of energies where phase shifts have been 
measured. 
We remind
the reader that potentials obtained by inversion are unique if and only if the
corresponding phase shifts are known for fixed angular momentum for all
energies, up to infinity, as required by Marchenko theorem \cite{chadan}. If there 
exist bound states, asymptotic normalization coefficients
for these states are necessary. Further, $\alp-\alp$ potentials are required 
to reproduce at least
qualitatively the known resonances in $^8$Be. These difficulties lead to a lot
of ambiguities in the proposed $\alp-\alp$ potentials. For the typical 
Ali-Bodmer potential the nonlocality translates into a strong 
dependence
of angular momentum in order to reproduce the repulsive effect of the redundant
states in the $L$=0 and $L$=2 partial waves. We shall use the $S$-state 
Ali-Bodmer potential as representative for the class
of phenomenological potentials with a soft inner repulsive core and a  weak long range
atractive component. 
There is however a more stringent requirement to be imposed on $\alp-\alp$
potential for $\alpha$ matter calculations.  In ref.\cite{cast79} the 
Euler-Lagrange equation for a Bose system within the HNC approximation 
was analized 
in detail and shown that a spherically symmetric solution which 
do not lead to a colapse of the system requires a potential with positive 
volume integral. 

The Ali-Bodmer potential the $\alp$ matter 
actually fails to saturate. In fact a very shallow minimum in the EOS at a high 
density is predicted.
Other schematic potentials (hard core) are trivially saturating at densities  and 
energies close to the nuclear matter 
saturation point ($\rho_\alp\approx$ 0.04 $\alp$/fm$^{3}$, and 
$E/N_\alp\approx$ -11-16 MeV).
However as we noted above, 
at such high densities the $\alpha$-condensate is almost completely depleted 
according to the variational approach. Somehow this dissapointing result 
is conflicting with what would one expect based on the manifestation of $\alp$ 
clustering in light real nuclei. The clusterization of $\alpha$ particles 
on the surface of nuclei at densities 
around half the central nuclear density, as revealed by $\alpha$-decay, 
$\alpha$-transfer reactions or the putative dilute three-alphas condensate in the 
Hoyle state of $^{12}$C are pointing to a higher stability of $\alpha$ 
matter at lower densities.

Since the potentials providing saturation at lower densities are highly 
schematic (infinite repulsive short-range interactions) we resort to a 
calculation of the bare $\alpha-\alpha$ interaction based on the 
double-folding method for two 
ions at energies around the barrier. As input we consider realistic densities 
of the $\alp$-particle and modern effective nucleon-nucleon interactions.

In the double-folding framework \cite{carst93} the interaction between two
alpha clusters is calculated as a convolution of a local two-body potential 
$v_{nn}$ and the single particle densities of the two clusters
\beq
v_{\alp\alp}(\bd{r})=
\int d\bd{r}_1  d\bd{r}_2 
\rho_{\alp}(\bd{r}_1)\rho_{\alp}(\bd{r}_2)v_{nn}(\rho,\bd{r}-\bd{r}_1+\bd{r}_2)
\label{dfold}
\eeq

The effective  $n-n$ interaction $v_{nn}$ is taken to  depend on the density $\rho$ of 
the nuclear matter where the two nucleons are embedded.
It should also consist of a density independent finite-range part with 
preferably two ranges such that a potential similar to the Ali-Bodmer is
obtained. We therefore choose the Gogny effective interaction.
Out of the three main parametrizations of this interaction, only the D1 \cite{DG80} and 
the most recent one D1N \cite{CGS08} are satisfying the
requirement of positive volume integral of the $\alp-\alp$ potential. 
The D1S parametrization \cite{bergerD1S}
leads to a potental with negative volume integral and is omitted.
.

In what follows we include only the direct part in the double-folding 
potential (\ref{dfold}). The knock-on nonlocal exchange component  
leads to a strongly attractive local component mainly due to the fact that the 
Perey-Saxon localization procedure is not reliable at the energies 
of interest considered in this paper.

We take a Gaussian nuclear matter distribution inside the $\alpha$-particle
\beq 
\rho_\alp(r)=4\left ( \frac{1}{\pi b^2}\right )^{3/2}e^{-r^2/b^2}
\label{rhoalf}
\eeq
with an oscillator parameter $b$ that corresponds to a root mean square (rms) 
1.58$\pm$0.004 fm resulting from a Glauber analysis of experimental interaction
cross sections \cite{AlK96}.

The direct effective $n-n$ force in the Gogny parametrization reads :
\beqa
v_{nn}^{\rm d}(\rho,\bd{r}_1-\bd{r}_2)&=&\oh\sum_{i=1}^2(4W_i+2 B_i-2 
H_i-M_i)e^{-|\bd{r}_1-\bd{r}_2|^2/\mu_i^2}\nn\\
&+&\frac{3}{2}t_3
\left [\rho\left(\oh(\bd{r}_{1}+\bd{r}_{2})\right)\right ]^{\gamma}
\delta(\bd{r}_1-\bd{r}_2)
\eeqa 
Inserting the gaussian density distribution (\ref{rhoalf}) in 
the double folding integral (\ref{dfold}) and using the Campi-Sprung 
prescription \cite{campi72} for the overlap density 
\beq
\rho(1,2)=\left (\rho_\alp(\bd{r}_1-\frac{1}{2}\bd{s})\rho_\alp(\bd{r}_2+\frac{1}{2}
\bd{s})\right )^{\frac{1}{2}},
\eeq
where $\bd{s}=\bd{r}_1+\bd{r}-\bd{r}_2$ is the $n-n$ separation in the
heavy-ion coordinate system \cite{carst93}, we obtain the $\alp-\alp$ potential,
\beqa
v_{\alp\alp}(r)&=&4\sum_{i=1}^2(4W_i+2 B_i-2H_i-M_i)\left 
(\frac{\mu_i^2}{\mu_i^2+2b^2}\right )^{3/2}e^{-{r^2}/{\mu_i^2+2b^2}}\nn\\
&+&\frac{3}{2}t_3\frac{4^{\gamma+2}}{(\gamma+2)^{3/2}(\sqrt{\pi}b)^{3(\gamma+1)}}
e^{-\frac{\gamma+2}{4b^2}{r^2}}
\eeqa
which is assumed to be the same in all partial waves.
\begin{figure}[h]
\center{
\epsfig{figure=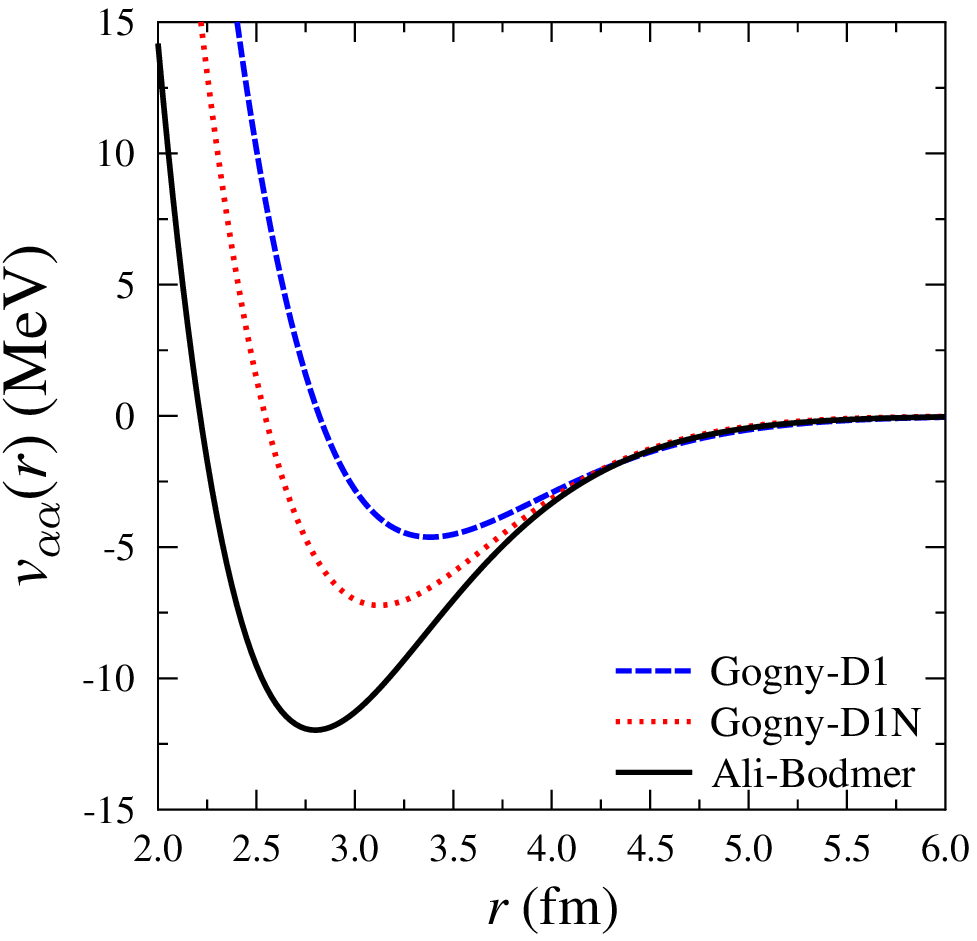,width=0.49\textwidth}
\epsfig{figure=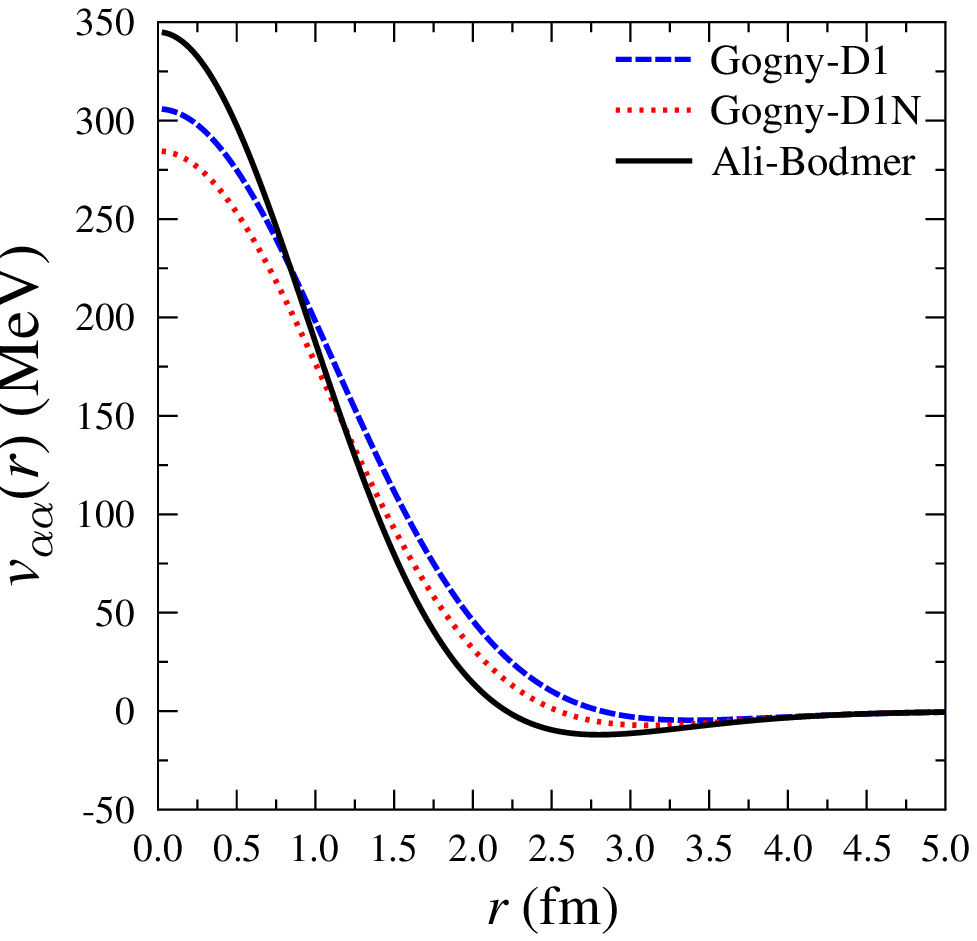,width=0.49\textwidth}}
\caption{
Different $\alpha-\alp$ potentials used in this paper. On the left panel 
we zoomed on the pocket region whereas on right the panel we display also the
soft-core.}
\label{Figpot}
\end{figure}
In the left panel of Fig.\ref{Figpot} we represent the three potentials on a 
magnified scale around the minimum. The two selected Gogny interactions display 
shalower pockets that are shifted to larger radii compared to the Ali-Bodmer potential. 
Before ending this section we remark that a fictious $^8$Be (Coulomb interaction 
is switched off) is slightly bounded by Ali-Bodmer $(J^{\pi}=0^+$, $E$=-1.63 MeV), 
loosely bounded by Gogny-D1N $(J^{\pi}=0^+$, $E$=-0.16 MeV) and not bounded by 
Gogny D1 interaction, in agreement with the patterns depicted in Fig. \ref{Figpot}. 
Levinson's theorem applied to the $\alpha-\alpha$ system \cite{neuda71} which 
predicts $\delta_0(E=0)=3\pi$, where $\delta_0$ is the scattering phase for 
the $L=0$ channel, is reasonably well satisfied by all potentials.

\section{Variational approach}
\label{sect3}

Within the variational approach to the description of a uniform system of 
spinless Bose particles at zero temperature interacting via a 
two-body potential $v_{\alpha\alpha}(r_{ij})$, the expectation value of 
the energy
\beq
E[\psi]=\frac{\langle\psi\mid H \mid\psi\rangle}{\langle\psi \mid \psi\rangle}
\eeq
can be calculated using the Bijl-Dingle-Jastrow (BDJ) trial wave 
function \cite{fe69}
 \beq
 \mid\psi \rangle=\prod_{i<j}f(r_{ij})
 \eeq
Above, $f(r)$ is a positive semidefinite function that is required to 
reflect exclusively the two-body correlations. Thus at short distances, 
where the potential is expected to be strongly repulsive, $f$ is small , 
whereas at large distances ($r\rightarrow\infty$), when the two-particle 
''decorrelate'', $f(r)\rightarrow$ 1.

The Jackson-Feenberg energy ${\cal{E}}_{\rm JF}$, measured relative to the rest energy of 
$N_\alpha$, $\alpha$-particles, is \cite{jf66,Cla66}
\beqa
{\cal{E}}_{\rm JF}&=&\oh\rho\int d\bd{r} 
g(r)\left [ v_{\alpha\alpha}(r)-\frac{\hbar^2}{2m_{\alpha}}\Delta\ln~f(r)\right ]
\label{energgs}
\eeqa
where $g(r)$ is the two-body radial distribution function (RDF).
This form of the energy has the advantage for bosons that terms including the 
three-body radial distribution function does not occur. 

The cluster expansion formalism heavily relies on 
the fact that $h=f^2-1$ is of short-range and consequently the integral $\omega=\int h(r)d\bd{r}$ is small 
compared to the volume $\Omega$ of the system 
($\omega=-\frac{1}{\rho}=\frac{\Omega}{N_\alpha}<<\Omega$)
\cite{ClaWe66}. Then $g$ can be evaluated by an expansion in 
powers of $\rho$ \cite{Mue70}. Including up to four-body diagrams
the expansion reads
\beqa
g(\bd{r}_{12})&=& f^2(\bd{r}_{12})\left \lbrace 1+\rho \int d\bd{r}_3 h(\bd{r}_{13})h(\bd{r}_{23})\right.\nn\\
&+& \frac{1}{2}\rho^2\int d\bd{r}_3\int d\bd{r}_4 \left [ 
2h(\bd{r}_{13})h(\bd{r}_{24})h(\bd{r}_{34})(1+2h(\bd{r}_{14}))\right.\nn\\
&+& \left.\left. h(\bd{r}_{13})h(\bd{r}_{23})h(\bd{r}_{14})h(\bd{r}_{24})
(1+h(\bd{r}_{34}) \right ] 
+{\cal O}(\rho^4)\right \rbrace
\label{rdfcluex}
\eeqa
The last term in the above formula include ring, diagonal, open and
elementary four-body diagrams. 

Retaining only the first term in the cluster expansion of $g$, 
the lowest-order Jackson-Feenberg functional is obtained upon 
substitution in the expression (\ref{energgs}) of ${\cal{E}}$ 
\beq
{\cal{E}}_2[f,\nabla f,\Delta f]=\oh \rho \int 
d\bd{r}\left\lbrace v_{\alpha\alpha}(r)f^2(r)
+\frac{\hbar^2}{2m}\left [(\nabla f(r))^2-f(r)\Delta f(r)\right ]\right\rbrace
\eeq
A practical method to obtain $f$ is provided by the Pandharipande-Bethe 
prescription \cite{PB73}. It consists in varying the two-body energy 
functional with respect to the one-parameter family of functions $f$ 
under the constraint of normalization $\rho\int[f^2-1]d\bd{r}=-1$. 
Thus, the variational problem is reduced to the eigenvalue problem
\beq
-\frac{\hbar^2}{m_\alpha}\left ( \frac{d^2 
f}{dr^2}+\frac{2}{r}\frac{df}{dr}\right )+\left ( v_{\alpha\alpha}-\lambda\right )f=0
\label{pbeq}
\eeq

\begin{figure}[t]
\center{
\epsfig{figure=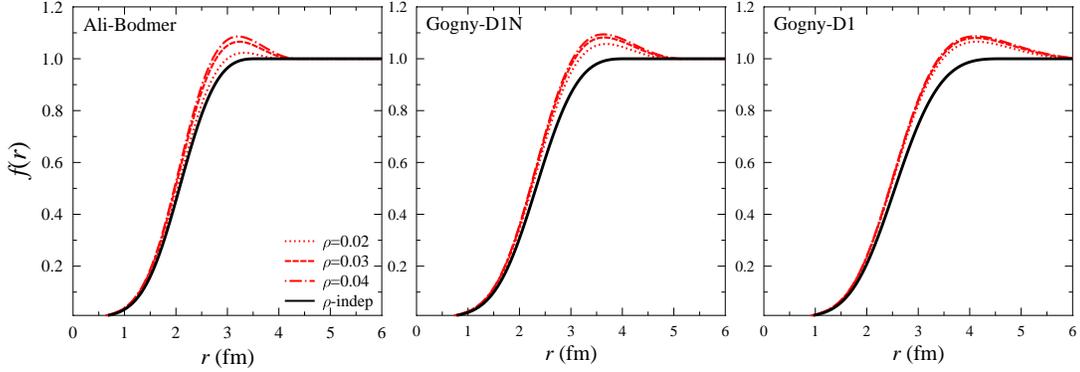,width=1.05\textwidth}
     }
\caption{Correlation function $f(r)$ resulting from solving eq.(\ref{pbeq}) for the two cases : 
a) dependent on density at three different densities and 
b) unique for all densities. Each figure contains these correlations functions for 
a given $\alp-\alp$ potential.}
\label{fcor}
\end{figure}

The CFN is subjected to natural boundary conditions that reflect the Schr\"odinger 
condition at the origin of a soft-core and the necessity to heal
to unity at a given distance $d$ 
\beq
u(r)\equiv r f(r)\stackrel{r\rightarrow 0}{\longrightarrow} 0, ~~~~f(r\geq d)=1, ~~~~f^\prime(d)=0
\eeq
For each density we varried the parameter $d$ untill the normalization
condition was fullfiled. In this way we obtained a solution that is density
dependent (since we take $d\sim\rho^{-1/3}$), and the overshoot 
(the peak  exceeding the unity)  increase with density as can be seen in 
Fig.\ref{fcor} for all three potentials.

Another recipe that we applied was to relax the normalization constraint
and match instead the potential at the right boundary with the 
eigenvalue $\lam=v_{\alp\alp}(d)$. This constraint is fullfilled for a unique 
$d$ regardless the value assigned to $\rho$.
The interpretation assigned by Pandharipande to the Lagrange multiplier $\lam$
is that of a contribution to the average field comming from the excluded 
particles. The corresponding CFN is then density independent and
it does not display any overshoot.
In both cases we select the lowest eigenvalue, corresponding to zero nodes 
of the CFN.

Naturally the validity of the cluster expansion is limited to low
densities if the truncation is performed after including the four-body 
correlations.
A method that allows the determination of $g$ and is more reliable
for higher densities is provided by the HNC approximation, where the diagrams
belonging to the nodal set ${\cal N}$ are summed up to infinity.
Thus, the RDF in HNC/0 approximation is defined as \cite{groene59}
\beq
g(r)=f^2(r)e^{N(r)}
\label{ghnc}
\eeq
The function $N(r)$ results from the iterative solution of a set of non-linear equations 
\beqa
X(r)&=&g(r)-N(r)-1\label{xdiag}\\
N(r)&=&\int d\bd{r}_1 d\bd{r}_2 X(r_1)(X(r_2)+N(r_2))\delta(\bd{r}_1+\bd{r}-\bd{r}_2)\nn\\
\label{nodalhnc}
\eeqa

Within the HNC/0 approximation elementary diagrams are neglected.
The first correction to this approximation is given by HNC/4 , where the
label 4 indicate that only the four-body elementary diagram is taken 
into account \cite{panschm77}. As a consequence expression (\ref{ghnc}) is 
modified, such that this diagram is added to the infinite number of nodal
diagrams :
\beq
g(r)=f(r)^2e^{N(r)+\epsilon_4(r)}
\label{ghnc4}
\eeq
where \cite{usmfripand82}
\beq
\epsilon_4(\bd{r}_{12})=\oh\rho^2
\int d\bd{r}_3 d\bd{r}_4G(\bd{r}_{13})G(\bd{r}_{23})G(\bd{r}_{14})
G(\bd{r}_{24})G(\bd{r}_{34})
\label{defeps4}
\eeq 
and $G(r)$ is a short-hand notation for $g(r)-1$. Thus, within HNC/4 approximation 
we obtain a fully selfconsistent solution of equations (\ref{xdiag})-(\ref{defeps4}).

\section{Ground-state energy of $\alpha$ matter}

The cluster expansion of the g.s energy results from inserting
the expansion in powers of densities of the RDF (eq.(\ref{rdfcluex})). The 
two-body and three-body terms are already given in the literature \cite{Mue70}. 
The four-body terms 
can be easily worked out employing the folding technique. Explicit
expressions of the four independent diagrams contributing to this term 
were derived in \cite{carst93}. 
On the other hand the g.s. energy in the HNC/0 approximation is obtained
by simply substituting into (\ref{energgs}) the RDF obtained from the 
iterative solution of eqs.(\ref{ghnc})-(\ref{nodalhnc})

\begin{figure}[t]
\center{
\epsfig{figure=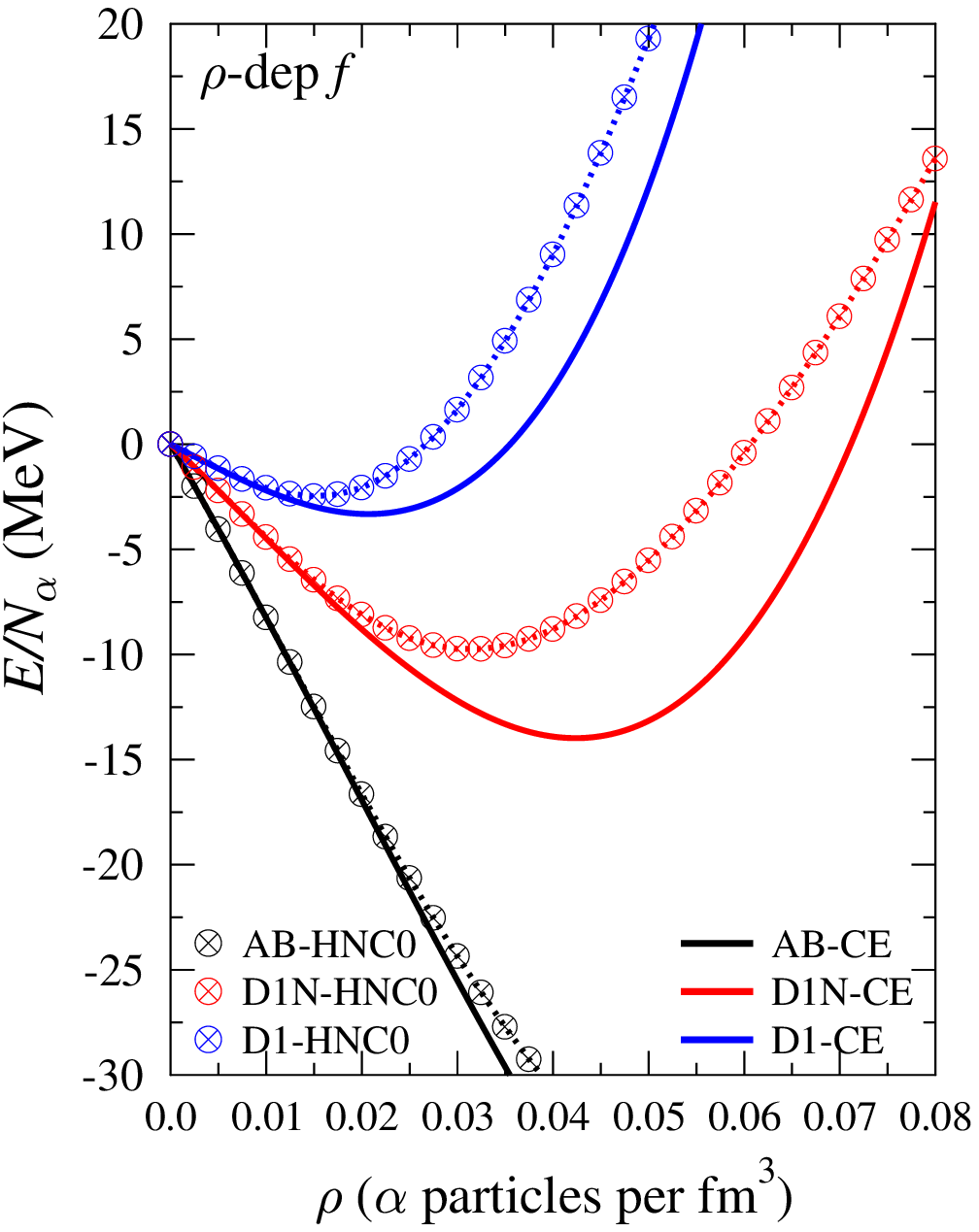,width=0.475\textwidth}
\epsfig{figure=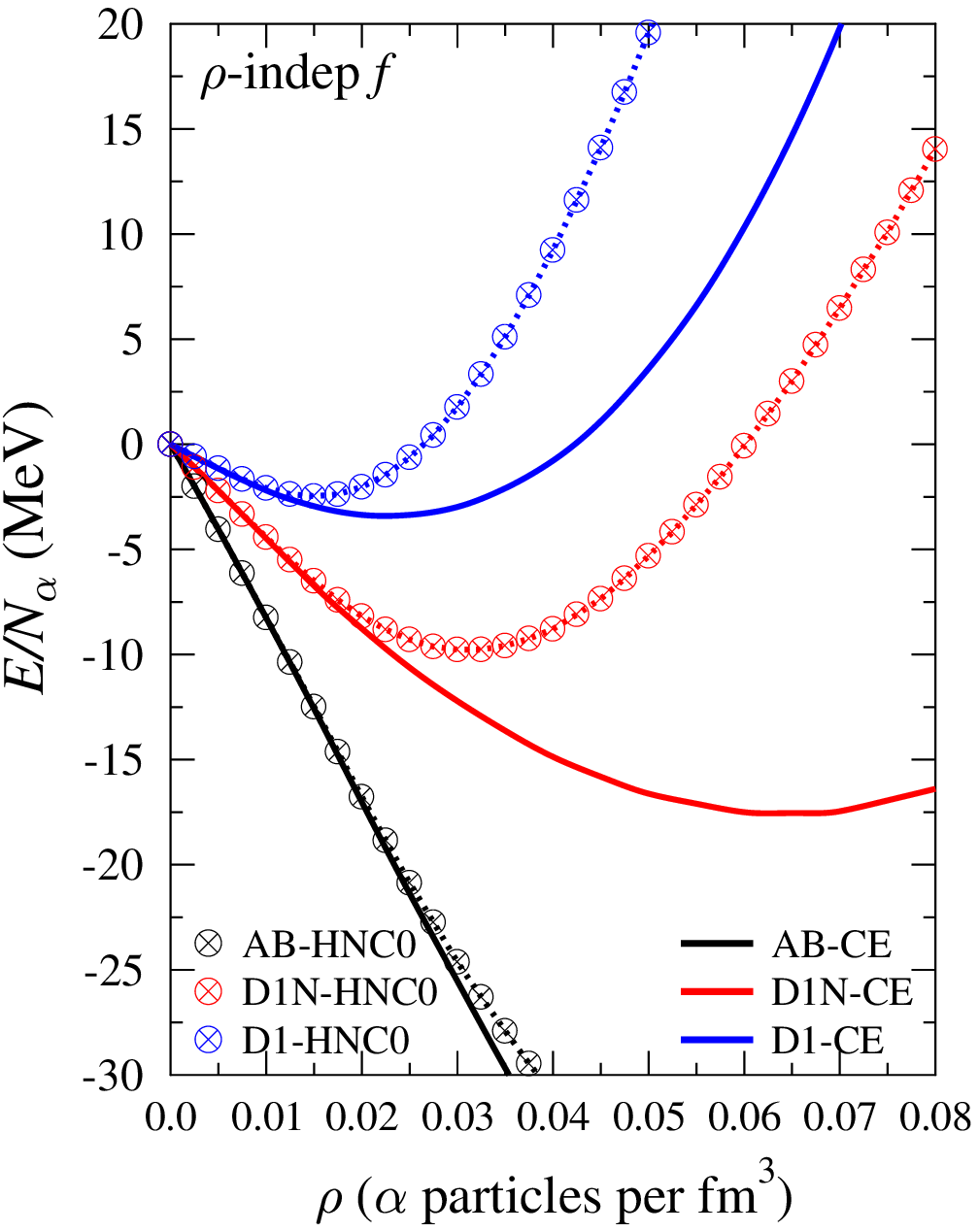,width=0.475\textwidth}
\caption{Total energy calculated with the cluster expansion (CE) taking into 
account up to four-body correlations contributions (solid lines) and with the 
HNC/0 method (circles) for the "$\rho$-independent" CFN considered in this paper.}
}
\label{eoscxhnc0}
\end{figure}

Fig.3 displays the energies for the three potentials
Ali-Bodmer, Gogny D1 and D1N calculated within the cluster expansion method
(solid curves) and with the HNC/0 method (circled crosses) 
for both CFN resulting 
from the Pandharipande-Bethe equation. 
Upon comparison with the $\alp-\alp$ potential we remark a similar 
pattern. It became obvious earlier that when we move from the most attractive 
potential (Ali-Bodmer) to the most repulsive  one (D1), the "molecular" pocket 
is shifted towards larger distances and shallower minima.
The saturation point of the $\alp$ matter EOS is considerably displaced to lower 
densities and instead of the overbinding manifest for the Ali-Bodmer potential 
we find a low-binding for the Gogny-D1 case. Another important feature of
the obtained EOS is that the HNC/0 calculations produce less bound $\alp$ matter 
with an obvious preference to saturate at lower densities.    
Although the cluster expansion formalism predicts softer EOS in the case
of the $\rho$-independent solution, the HNC/0 results are almost identical
for the two different choices of the CFN, a fact that suggests  
that most likely both are close to the optimal solution. As noted 
in \cite{zabol81} for the Bose homework problem, the fact that 
different looking correlation functions may give extremely close energies, is a
general feature of the variational method.

The cluster expansion and HNC methods are similar over a larger range for the Ali-Bodmer potential, i.e. up to $\rho\approx$ 0.0225 $\alp$ per fm$^{3}$ whereas for the Gogny-D1 potential the interval is limited to $\rho\le$ 0.0125 $\alp$ per fm$^{3}$.

\begin{figure}[t]
\center{
\epsfig{figure=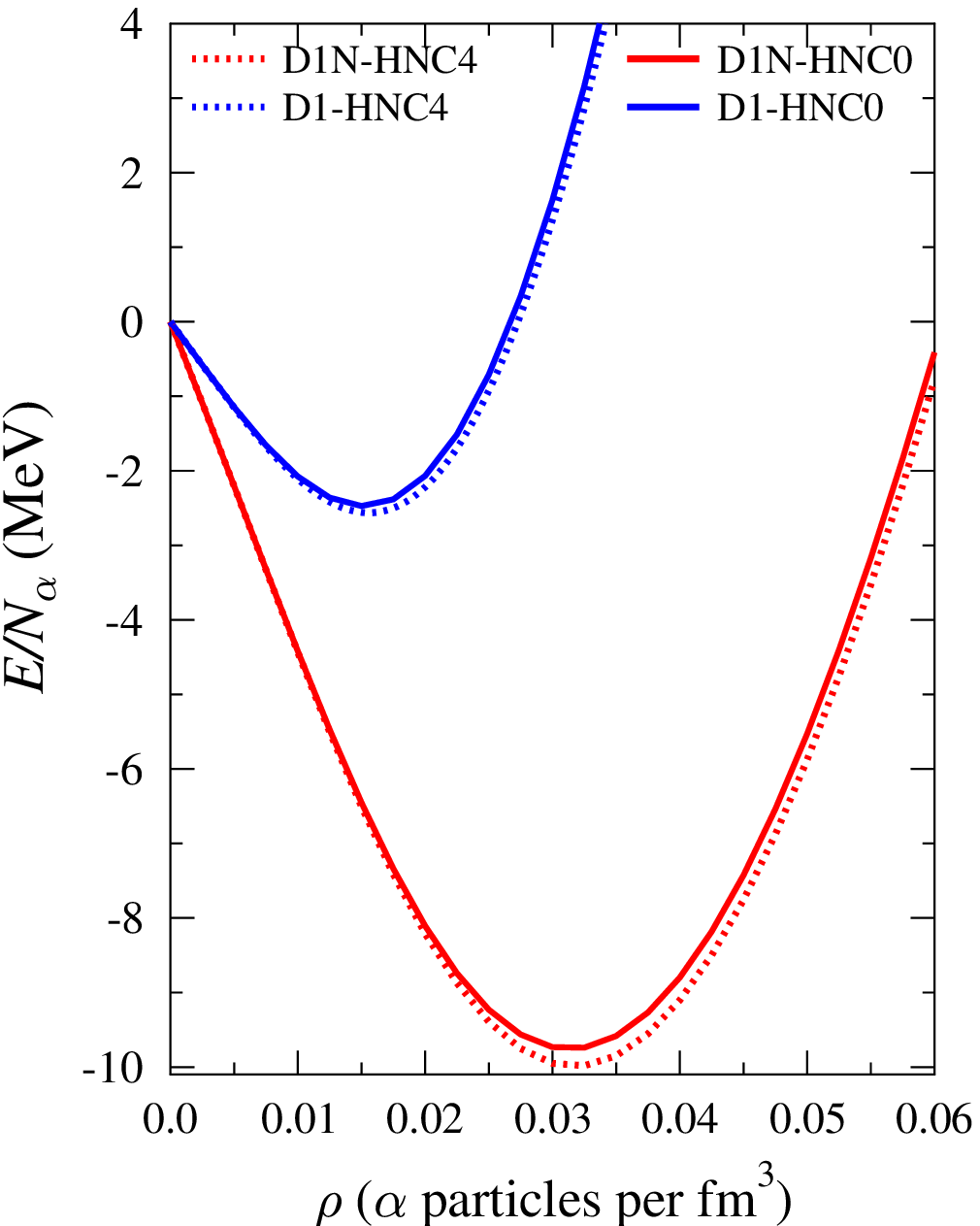,width=0.475\textwidth}
\epsfig{figure=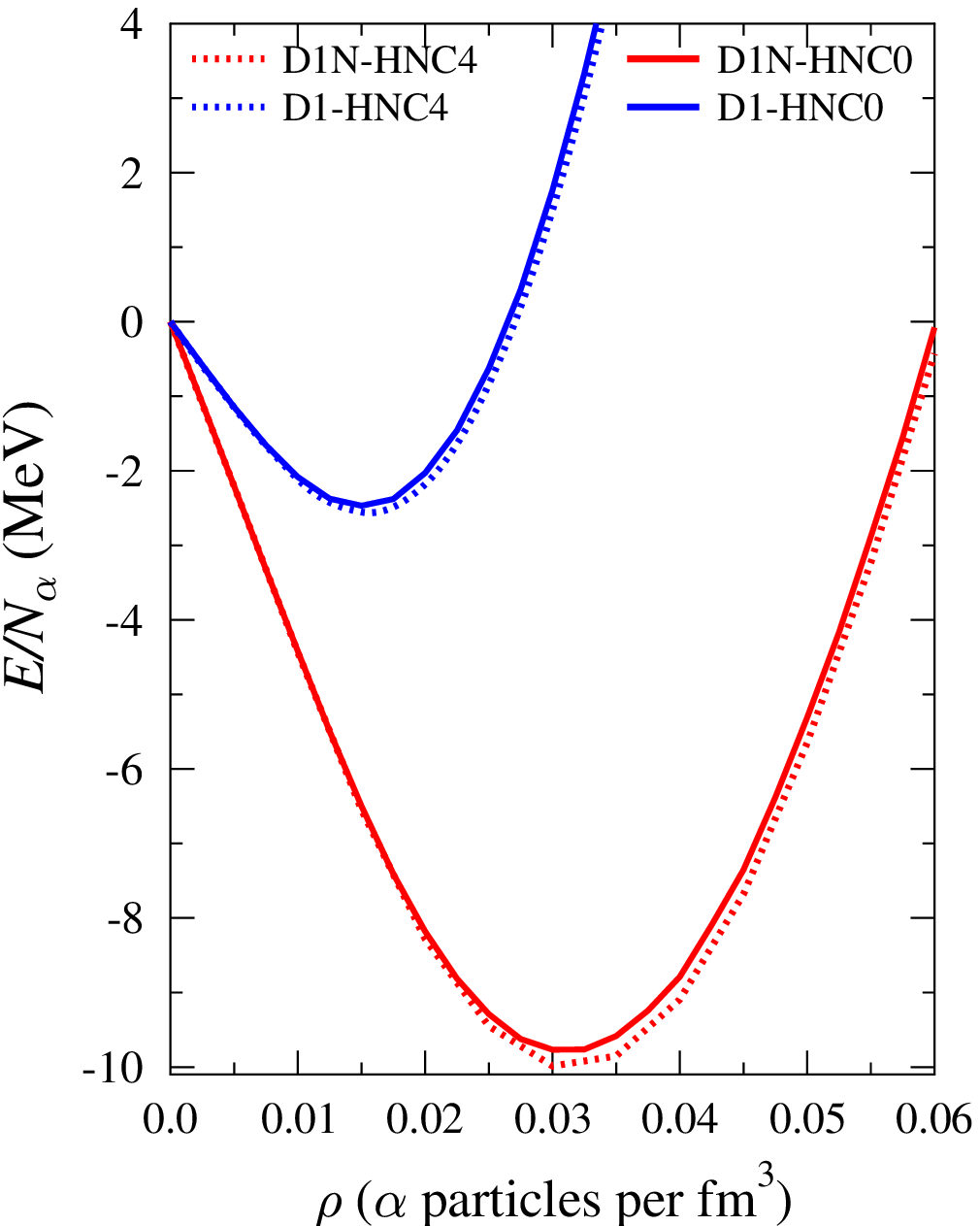,width=0.475\textwidth}
\caption{Comparison between the EOS predicted with HNC/0 and HNC/4 for 
D1 and D1N potentials for both CFN selections.}
}
\label{eos_hnc0vshnc4}
\end{figure}

Previously HNC/4 calculations have been reported for the $^4$He quantum 
liquid \cite{usmfripand82,smith79} but to our knowledge there is no similar 
calculation for the g.s. of $\alpha$ matter. In Fig.\ref{eos_hnc0vshnc4}
we compare the EOS predicted in the two HNC approximations using the 
Gogny potentials. Apparently the contribution of the $\epsilon_4$ diagram 
results only in a small lowering of the saturation energy.

\begin{figure}[h]
\center{
\epsfig{figure=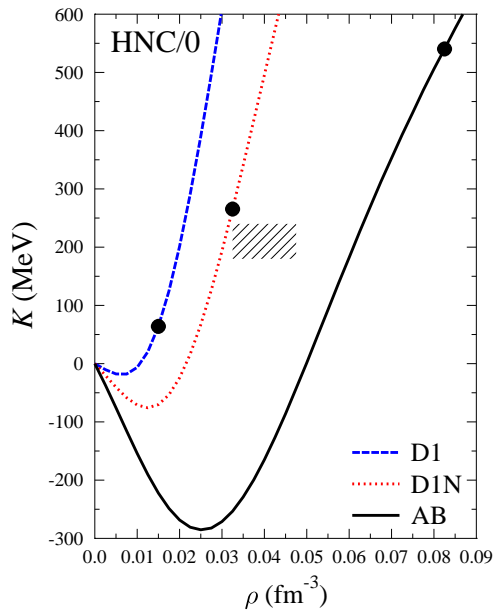,width=0.475\textwidth}
\caption{Incompressibility of $\alpha$ matter within HNC/0 for the 
three interactions Ali-Bodmer, D1 and D1N. The full circles specifies 
the 
saturation points and the dashed region the accepted range of symmetric nuclear matter saturation values.}
}
\label{compreshnc0}
\end{figure}

From the EOS derived within the HNC/0 formalism we extracted the 
corresponding incompressibilities as a function of density (see Fig.\ref{compreshnc0}). Although 
the EOS for the Gogny-D1 interaction has the largest curvature, since 
the saturation takes place at a rather small density ($\rho_{\rm sat}
\approx$ 0.015 $\alp$ per fm$^{3}$), 
the incompressibility is also small , $K\approx$ 64 MeV. Instead the  
Gogny-D1N interaction which predicts a saturation point at approximately 
$\rho_{\rm sat}\approx$ 0.0325 $\alp$ per fm$^{3}$ and binding energy -16.8 MeV, provided we add the -7.07 MeV binding energy per nucleon of 
the $\alp$ particle, has a corresponding value of $K$ close to the generally accepted range of standard nuclear matter values, i.e. $K\approx$ 265.4 MeV. 
We note that within the dynamical lattice model of $\alp$ matter with a 
Brink-Boeker I effective interaction by Tohsaki, a volume energy of 
-16.76 MeV and incompressibility $K\approx 181$ MeV are predicted \cite{tohs96}.

From the inspection of Fig.\ref{compreshnc0} we conclude that the 
interval where $\alp$ matter develops thermodynamical instability ($K<0$) 
is ranging over dilute densities for the Gogny D1 and D1N potentials, 
whereas for the Ali-Bodmer interaction also large values of the densities are 
affected by 
instability.    

In our opinion the domain of thermodynamical instability provides a further
ground to discard the Ali-Bodmer potential since it predicts a highly 
unstable $\alp$ matter at densities where $\alp$ clusterization is 
believed to be important. The Gogny D1 potential has saturation 
properties very close to the symmetric nuclear matter as can be noticed on 
Fig.\ref{compreshnc0}. On the other hand although the Gogny-D1 interaction 
predicts a much softer EOS and a low binding energy, it displays instead
stability at densities not far from the Mott density \cite{roepke99}.
We therefore expect a realistic $\alp$ matter EOS to develope a saturation point 
approximately in the domain bounded by the saturation points of the D1 and D1N EOS.

\section{Reduction of the Condensate Fraction}

Another quantity  of interest for the 
$\alpha$ matter ground state is the fraction of particles $n_{\scriptscriptstyle{C}}$ filling the 
zero momentum state (BEC) . 
Using a diagramatic expansion, Ristig and Clark showed that 
$n_{\scriptscriptstyle{C}}$ can be represented by irreducible 
$n$-body cluster diagrams $\Delta Q^{(n)}$, with two external 
points and $n-2$ internal field points \cite{ris76} :
\beq
n_{\scriptscriptstyle{C}}=e^{\Delta Q^{(1)}+\Delta Q^{(2)}+\Delta Q^{(3)}+\ldots}
\eeq
For the first three terms closed forms are available. Introducing the 
notation $\zeta(r)=f(r)-1$ and making use again of the compact 
representation of integrals via the convolution operation, the 
corresponding expressions are
\beq
\Delta Q^{(1)}=\rho\int d\bd{r}\zeta^2(r)
\eeq
\beq
\Delta Q^{(2)}=\rho^2\int d\bd{r}\left\{\zeta(\bd{r})(\zeta*G)(\bd{r})-\oh
h(\bd{r})(h*G)(\bd{r})\right\}
\eeq
\beqa
\Delta Q^{(3)}&=&\frac{1}{6}\rho^3\int d\bd{r}_2d\bd{r}_3d\bd{r}_4(2\zeta(\bd{r}_{12})\zeta(\bd{r}_{13})\zeta(\bd{r}_{14})-
h(\bd{r}_{12})h(\bd{r}_{13})h(\bd{r}_{14}))\nn\\
 &~& \times[g_3(\bd{r}_2,\bd{r}_3,\bd{r}_4)-G(\bd{r}_{23})-G(\bd{r}_{43})-G(\bd{r}_{24})-1]
\eeqa
The calculation of the third-order contribution is greatly simplified by the fact 
that at the HNC/0 level, the tree-particle radial distribution function is 
given by the Kirkwood superposition approximation
\beq
g_3(\bd{r}_2,\bd{r}_3,\bd{r}_4)=g(\bd{r}_{23})g(\bd{r}_{43})g(\bd{r}_{24})
\eeq
Note that the three-body RDF is large only when all three alphas are 
close to each other (within distances of order $\rho^{-{1}/{3}}$).
Details on the calculation of this last term will pe presented 
elsewhere. 

\begin{figure}[t]
\center{
\epsfig{figure=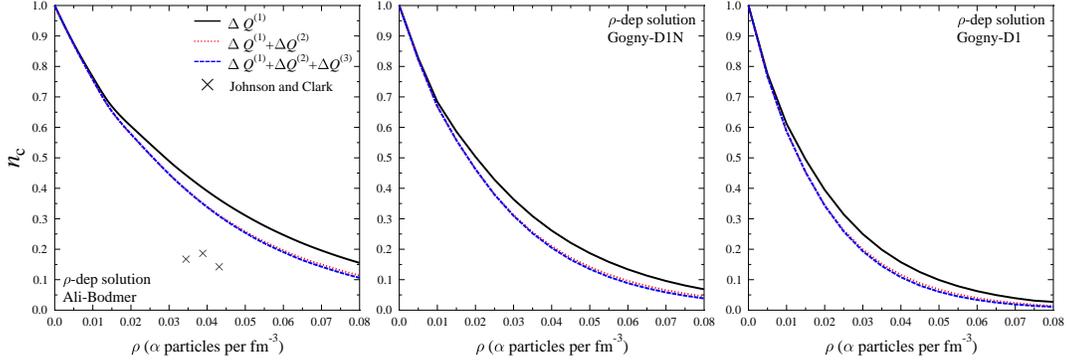,width=1.05\textwidth}
\caption{Reduction of the condensate fraction due to the contribution of 
higher order terms in the cluster expansion of the function $Q$
for the three potentials used in this work.}}
\label{redcondfrac}
\end{figure}

The exponential reduction of the condensate fraction is represented in 
Fig. \ref{redcondfrac} for the case when the second-order and the second 
plus third-order terms are added to the lowest order contribution for the 
three potentials used in this work. We perform an exemplification using 
only the density-dependent CFN. In all cases the third-order diagram is only
slightly reducing $n_C$ whereas the contribution of $\Delta Q^{(2)}$
is important beyond a certain density that is approximately 0.01 
$\alp$/fm$^{3}$.
As expected, in the case of the Ali-Bodmer potential the depletion  
of the condensate state with increasing density is smaller that in the case of 
Gogny potentials. Thus, whereas in the Ali-Bodmer case at 
saturation densities of nuclear matter ($\approx$ 0.04 $\alp$/fm$^{3}$)
the depletion is around 35$\%$, for the repulsive Gogny-D1 potential
only a tenth of alphas are left in the condensate. 
On the other hand the HNC calculations employing Gogny potentials are 
predicting saturation at densities 
$\rho\approx$0.015 $\alp$ per fm$^{3}$ (D1) and  
$\rho\approx$0.0325 $\alp$ per fm$^{3}$ (D1N). 
As can be inferred from Fig.\ref{redcondfrac} the condensate at this density 
is still well populated, i.e. 55$\%$ (D1) and 30$\%$ (D1N).

\section{Conclusions and Outlook}

The properties of cold $\alpha$ matter have been analyzed using the framework
of the variational theory. Compared to previous studies, performed
mainly by Clark and collaborators, we introduce two new $\alp-\alp$
potentials that are based on the realistic $\alpha$-particle densities and 
effective Gogny $n-n$ interactions widely used in modern nuclear structure 
studies. 
Hypernetted chain calculations including the elementary 
4-body diagram are providing slight corrections to the HNC/0 approximation.
For the low-density regime the cluster expansion
method is providing results similar to the HNC, which means that predicitions
of the $\alp$ matter g.s. properties in this case can be safely and rapidly 
obtained taking into account at most four-body correlations.         

To our knowledge for the first time in the literature a detailed investigation 
of the condensate fraction for $\alpha$ matter was performed. The reduction 
of condensate fraction is enhanced once we take into account higher-order 
contributions in the cluster expansion of $Q$. For the Ali-Bodmer
case the reduction is temperated by the more attractive character of the potential. 

We inferred that only the Gogny-D1 based potential is predicting a saturation 
of alpha matter close to the Mott density. Moreover at this density
there is a significant condensate fraction. On the other hand the 
Gogny-D1N EOS 
displays saturation properties (density, energy and incompressibility) similar
to the symmetric nuclear matter EOS. For this reason we are confident that a
realistic saturation point of the $\alpha$ matter should be located approximately
in the range between 0.015 - 0.0325 $\alp$ per fm$^{3}$. 
 


As mentioned in the introductory section the main
factor that triggered this work was the renewed interest in the literature 
regarding the properties of the $\alpha$-condensate. We showed that 
compared to the dilute approximation adopted in ref.\cite{Sch07}, the 
inclusion of higher order terms in the cluster expansion, produce a steeper 
depletion of the condensate, which is even more accentuated for the repulsive 
Gogny potentials.

The present work was also motivated by modern theories of low-density
nuclear matter composed of neutrons, protons and alpha particles near  the
neutrinosphere during supernovae core-collapse (see \cite{horow07}).
In this context it would of interest to extend the present calculations to the case of non-zero temperature and inhomogenous distribution of $\alp$ matter.

\section{Acknowledgement}

We are gratefull to M. Rizea for many usefull discussions. 
This work received partly support from CNCSIS Romania, 
under Programme PN-II-PCE-2007-1, contracts No.49 and No.258.


\end{document}